# Color continuity along the journey from ideas to objects

Ján Morovič

*HP Inc., Barcelona, Catalonia, Spain*
https://www8.hp.com/uk/en/commercial-printers/

**1. Introduction**

Beatrice Warde's maxim that "type well used is invisible as type" can be applied to many other domains, including that of color communication. The ultimate goal here is for there to be no need whatsoever to pay attention to color and for it to "just work". Instead of having to care about how color is communicated, how it is represented, how it is measured, how it is managed, how it is kept consistent and how it relates across different contexts, technologies and applications, it should "just work". The ultimate objective is for color communication to happen invisibly, without the need for intervention, but for it to happen as expected – consistently with different and multiple possible expectations under different circumstances.

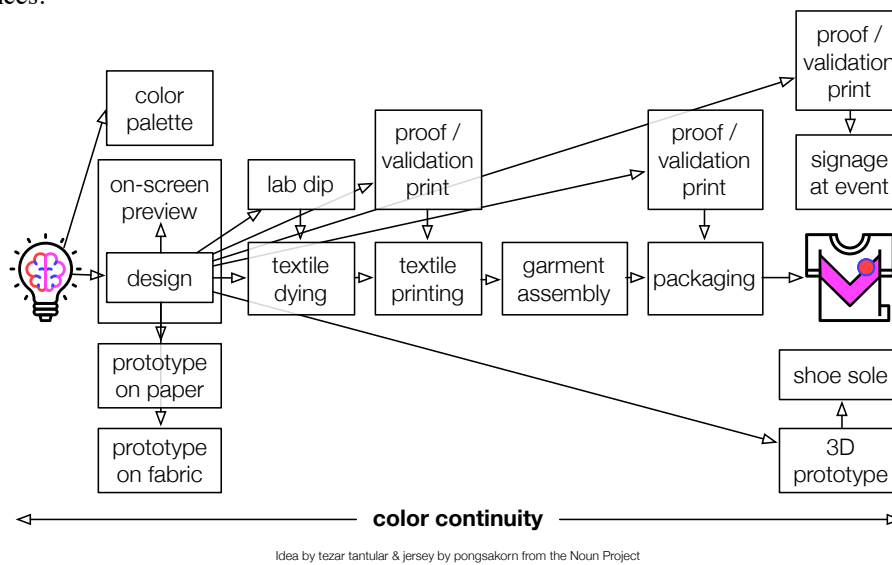

**Figure 1:** From ideas to objects: a bicycle race jersey case study.

Taking the example of a team jersey for a bicycle race, it is possible to trace a journey from a creative team's minds to the jersey being delivered to its wearer ahead of an event (Fig. 1). The starting point would involve the challenges of a first materialization of ideas, where the definition of a color palette (using swatches from color books, found objects, …), on-screen previews, coupled with first printed prototypes on paper and eventually prints on fabric, would be part of an iterative process of arriving at a design. The color focus here would be on expressing a feeling, a mood, a style, a level of comfort or dynamism, etc. and representations of the design would serve the purpose of validating choices against such broad concepts. A prototype on fabric may also be produced to gage how the design drapes, stretches, fits. This process may also include the incorporation of pre-existing designs, such as team or sponsor logos, with their own color specifications. A combination of textile dying, where accurate matching to the initial design under multiple light sources would be key, and printing on pre-dyed textiles may then follow, where color matching between different materials and to tight tolerances would be the objective. Here the challenges of maintaining color consistency during different production processes (often also across geographies and technologies), and of configuring them for matching pre-defined targets and thresholds under a variety of viewing conditions, would be major constraints.

A garment prepared in such a way would then be assembled and packaged, where the packaging too would be the result of a parallel journey from idea to object and where color continuity with the garment itself would be needed. At the race there may then also be signage (posters, banners, vehicle or helmet wraps, …) as well as other objects whose colors relate to that of the jersey (backpacks, bags, lanyards, pens, personalized labels, …), including 3D-printed ones like shoe soles, which again would have had their own idea-to-object color journeys that would share a common requirement of color continuity.

All of the above may sound either like science-fiction (since the number of players and elements in these processes is very large, the connections among them are incomplete and managing them is challenging) or trivial (since we see examples of it being done reasonably or even very successfully today). Instead, it can be thought of as representing a space of performance versus complexity and time where current practice is at a certain point that today requires significant expertise and iteration. At the same time, there is a movement towards a much more closely integrated and streamlined way of creating and producing that will allow for more energy being directed at adding value than at having to manage and coordinate processes. The next section will present examples of recent developments in color printing that are along the path towards making the handling of color transparent, followed by a section that sets out some key challenges in color communication whose resolution lies ahead for the next 10 years or more.

## 2. Recent advances in color for printing

As examples of how printing is changing in a way that moves towards enabling a more transparent handling of color, this section will introduce examples from HP for the sake of showing tangible cases of color and imaging search having made it from discovery/invention, through product development and into productization. At the same time, it is important to acknowledge that all players in the printing ecosystem are making advances towards greater streamlining and integration and could give their own examples of how color printing has been advanced over the last years. Turning to HP though, the following are four prominent instances of such evolution and even revolution.

First, there is the $\Delta E_{NS}$ color difference equation [1], developed in collaboration with Zhejiang University in response to the need for more accurately predicting the level of perceived color difference between two stimuli viewed with no separation (NS). While color difference equations such as $\Delta E2000$ and $\Delta E(CMC)$ are derived from observations about viewing stimuli with a clear gap between them and against a uniform background, many printed objects involve parts made at different times or coming from different locations within a substrate viewed side-by-side with no separation (Fig. 2a). Tiling for billboards or wallcoverings, packaging with folded carton or printing of garments assembled out of multiple pieces all share this characteristic. Having such a new metric for color difference then becomes an enabler for handling various color processes (e.g., calibration, verification) more reliably when their output is used in gapless viewing scenarios and where previous metrics left a disconnect between their predictions and observer experiences that required greater manual intervention.

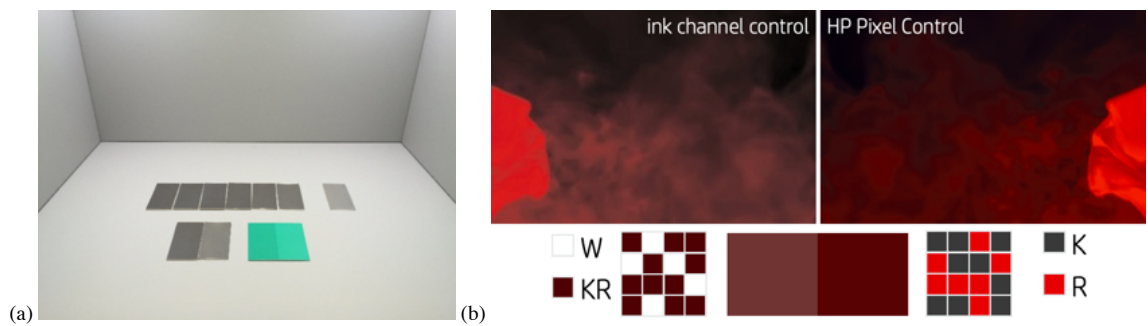

**Figure 2:** (a) experimental setup for deriving $\Delta E_{NS}$ color difference equations, (b) basic principle of HP Pixel Control imaging pipeline where at-pixel states, instead of only ink quantities, are controlled directly. This allows for richer dark colors resulting from side-by-side printing of black (K) and red (R) ink, which, without pixel control can be just overprinted, resulting just in a dull color.

Second, there is a fundamental re-thinking and re-engineering of how printing is controlled in the form of HP Pixel Control (internally codenamed HANS) [2, 3]. Looking at a print, it can be seen that its colors depend on the relative proportions with which the various combinations of a substrate and a set of inks are used at small scale. These substrate-ink combinations – called Neugebauer Primaries (NPs) – were first identified by Hans Neugebauer in the 1930s as the basic building blocks of predicting print color, which results from their convex combination:

$$T(NPac_C) = \sum_{i=1}^{k^n}(w_{Ci} * T(NP_i)) \quad (1)$$

where pattern C is formed by combining some of a printing system's NPs and can be characterized by its NP area coverage (NPac) vector – $NPac_C$, $k$ is the number of colorant levels per colorant per pixel, $n$ is the number of colorants, $\sum_{i=1}^{k^n} w_{Ci} = 1$ (i.e., the weights are convex), $NP_i$ is the $i$-the Neugebauer Primary, and $T()$ is color (e.g., the NPs' Yule-Nielsen corrected colorimetry).

With HP Pixel Control, NPs become the control domain for printing. Instead of starting with a translation of color to colorant amounts (e.g., ink, toner, …), as has been the case with print control since the late 19[th] century, HP Pixel Control determines how much relative area to cover with each of a system's possible halftone pixel states – the Neugebauer Primaries. This unlocks a vast variety of ways in which printable patterns can be formed, which in turn gives access to new printable colors, more efficient colorant use and higher image quality, to name but a few (Fig. 2b). For the case of a 3 ink printer, the difference between its 3 inks (e.g., CMY) and the 8 Neugebauer Primaries that correspond to it (S – the blank substrate, C, M, Y, CM, CY, MY, CMY) is already significant, but when it comes to a printer with 14 ink channels, each of which can specify up to 3 drops per ink per pixel, the result is the ability to form patterns using over 268 million pixel types instead of 14 inks. Finally, HP Pixel Control also exercises control in terms of pixel area coverages instead of ink amount, which is the same domain in which the human visual system integrates content below the spatial acuity threshold. In the first printer with a fully-digital HP Pixel Control pipeline, the HP DesignJet Z9[+], this leads to more natural looking prints, with smoother color transitions and gradations and with greater detail, all of which also makes color management and predictability perform better and enable more effective automation.

Third, there is the novel use of embedded color sensors for consistent color production. A recent example here is the on-board spectrophotometer in the HP Stitch S Series dye sublimation textile printers (Fig. 3a), which allows for a particularly efficient way of achieving consistent color output. Most of the fabrics printed using dye sublimation involve a two-step process: a print is first made on a transfer paper and the inked pattern it holds is then transferred to a fabric using a calendering machine, where under heat and pressure the ink is sublimated (becomes gaseous) and interacts with a fabric's fibers. From a color configuration and calibration perspective, such a two-stage workflow presents significant challenges and the way it is dealt with typically involves several printing-sublimation-measurement stages, which can make color setup last hours. While the appearance of the print on the transfer paper is not that of the final fabric, having an on-board spectrophotometer in the printer means that the printed color can be kept to a very high level of color consistency in a very quick and operator-free way. A calibration chart can be printed, measured in situ on paper and calibration look-up tables can be updated in a matter of minutes, saving both time and materials and again allowing for greater automation and unattendedness.

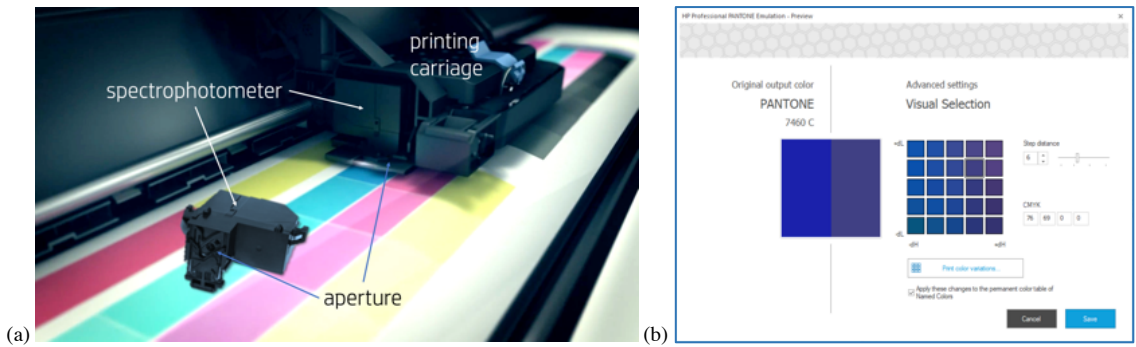

**Figure 3:** (a) Spectrophotometer on-board a printing carriage, (b) HP Pantone refinement interface for visual adjustment.

Fourth, there are advances in making complex color processes simple and streamlined and thereby unlocking their benefits for a broader user base. An example here is the creation of ICC output color profiles for the HP Stitch S Series printers. Instead of exposing all of the many and complex parameters and choices of color profiling to a user, and thereby requiring that they have the necessary expertise and experience to set them, HP Easy Profiling comes with pre-configured settings that will work well for a large proportion of applications and workflows. All that a user then has to do is trigger the profiling process, indicate for which substrate and printer it is to be run and name the resulting profile. Everything else is pre-set and executed without the need for expertise by the end user. As a result, a high level of color accuracy is made accessible to users without color expertise and the process is made quicker even for more seasoned operators. The same printer also comes with a solution to make the visual adjustment of spot color emulation faster and easier for colors outside a printing system's color gamut. Even though there are already solutions for directly identifying the visually closest color for an out–of–gamut spot color, such as the algorithm used in HP Professional PANTONE Emulation [4], there is an element of preference at play. Here a user's task can be made simpler by presenting them with the most likely alternatives to the visually closest one. This is done by sampling not the device color space of a printing system (e.g., CMYK or RGB), but by presenting alternatives from the surface of the output color gamut that vary in hue and in lightness around the visually closest choice. Offering control over the third color dimension, chroma, is not meaningful here, since it could only lead to less chromatic choices at a given lightness-hue combination, and therefore to colors that are even further away from the target.

## 3. Key challenges for the next 10 years of color communication

Looking ahead, four areas for future work stand out in particular. First, that of fundamentals, of being able to predict with greater accuracy and breadth of coverage richer and more varied scenarios the color resulting from a variety of processes including printing and post-processing (e.g., celendering, laminating, varnishing) and using materials like fluorescent colorants and non-white substrates. Second, that of closing the loop, of process control that involves efficient color verification and the application of vision systems capable not only of detecting spatial detects but also color changes to a high degree of accuracy and consistency. Third, that of end-to-end workflows, of connecting, guiding and auto-configuring choices, resources and processes from initial idea to final product. Here an incorporation of production constraints into the design process is a particularly challenging and promising idea. Being able to easily anticipate what a design will look like in production and getting guidance about how to adjust an initial approach so that it conveys its creators' intent would help greater predictability and allow for shortening the end-to-end journey. Working with multiple materials in a project and using multiple imaging systems to produce the same color output are also important challenges here. Fourth, that of optimization for a variety of attributes from color and image quality, via consistency to cost and resilience to production variation are all central to making color printing and other forms of color communication become more widely used in a greater variety of contexts. To make significant advances in these and other, yet to be identified, areas will require both fundamental color and imaging work and greater collaboration with other disciplines. While color communication is as old as human civilization, in many ways it is only just getting started.

**Acknowledgements:** I would like to thank the broader color and imaging community and especially my colleagues at HP Inc.: Africa Real, Albert Serra, Alessandro Beltrami, David Duran, Francesc Costa, Hector Gomez, Javier Maestro, Jordi Arnabat, Jordi Vilar, Konstantinos Kontonikolas, Pere Gasparin, Peter Morovič, Sergio Etchebehere, Tanausu Ramirez, Xavier Fariña.